\documentclass[aps,prb,twocolumn,showpacs,floatfix,preprintnumbers,superscriptaddress,amsmath,amssymb,footinbib]{revtex4-1}

\usepackage{graphicx}
\usepackage{dcolumn}
\usepackage{bm}

\newcommand{\sbr}[1]{\left[#1\right]}

\begin{document}

\preprint{ptRel v3.0 arXiv}

\title{Relativistic peculiarities at stepped surfaces: \\surprising energetics and unexpected diffusion patterns}

\author{O. O. Brovko}
\affiliation{Max-Planck-Institut f\"{u}r Mikrostrukturphysik, Weinberg 2, D06120 Halle, Germany}
\author{N. N. Negulyaev}
\affiliation{Fachbereich Physik, Martin-Luther-Universit\"at, Halle-Wittenberg, F.-Bach-Platz 6, D06099 Halle, Germany}
\author{V. S. Stepanyuk}
\affiliation{Max-Planck-Institut f\"{u}r Mikrostrukturphysik, Weinberg 2, D06120 Halle, Germany}

\date{28 September 2010}

\begin{abstract}
    We revive intriguing, yet still unexplained, experimental results of Ehrlich and coworkers [Phys.~Rev.~Lett. \textbf{77}, 1334 (1996) and Phys.~Rev.~Lett.~\textbf{67}, 2509 (1991)] who have observed, that \emph{5d} adatoms distributed on (111) surface islands of \emph{5d} metals favor the adsorption at the cluster's edge rather than at the cluster's interior, which lies in contrast with the behavior of \emph{4d} and \emph{3d} elements. Our state of the art \emph{ab initio} calculations demonstrate that such behavior is a direct consequence of the relativity of \emph{5d} metals.
\end{abstract}


\maketitle

    The recent surge of attention to nanoscale noble metal systems, serving as a base for nanostructured materials and paving the road to novel electronic devices and nanocatalytic agents, has reignited the community's interest in numerous peculiar properties of \emph{5d} elements. Among the abundance of experiments done over the past two decades on \emph{5d} metals some still remain not completely understood despite fascinating and potentially fundamental results that have been obtained.

\begin{figure}
		\includegraphics{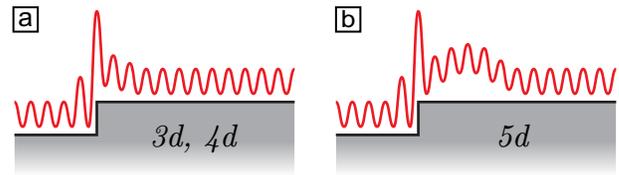}
        \caption{\label{fig:1} (color online) Schematics of the potential-energy landscape (a) which govern diffusion in \emph{3d} and \emph{4d} metallic systems \cite{Bromann1995,*Morgenstern1998,*Prieto2000,Mo2008} and a landscape (b) which corresponds to the diffusion behavior observed in experiments on Pt and Ir \cite{Golzhauser1996,*Wang1991,*Wang1993}.}
	\end{figure}

    In the present paper we revive the results of Ehrlich and coworkers \cite{Golzhauser1996,*Wang1991,*Wang1993} who have observed, that the energetics and diffusion patterns of Pt on Pt(111) and Ir or W on Ir(111) display quite surprising features. In a classical adsorption picture, known to be true for \emph{3d} and \emph{4d} metals \cite{Ehrlich1966,*Schwoebel1966,Mo2008}, single adatoms diffusing on stepped or island-covered surfaces at low temperatures tend to avoid areas close to descending step edges. Such repulsion is usually attributed to the consequences of the electron density smearing at step edges, known as the Smoluchowski effect \cite{Smoluchowski1941}. A sharp step in the atomic potential causes electrons to redistribute (flow down from the step) so as to smoothen out the electron density profile. The resulting electronic depletion in the vicinity of a descending step edge makes adsorption there unfavorable, thus introducing a barrier (a few \AA~wide) for diffusing adatoms. The depletion is strongest directly at the step so that diffusing atoms, once able to overcome the edge repulsion, tend to jump down onto the lower terrace or incorporate into the step \cite{Bromann1995,*Morgenstern1998,*Prieto2000,Mo2008}. Such diffusion behavior can be described by a potential energy profile schematically presented in Fig.~\ref{fig:1}a.

    The first surprising feature found in experiments by Ehrlich and coworkers \cite{Golzhauser1996,*Wang1991,*Wang1993} was that while the step repulsion barrier was still observed for the diffusion of Pt on Pt(111) and Ir or W on Ir(111) its width was significantly increased (up to 10~\AA). Yet the most surprising finding, was the fact that once the temperature was raised high enough for the atoms to overcome the barrier introduced by the electronic depletion at the cluster's edge, the atoms did neither jump onto the lower terrace nor incorporate into the step but tended to migrate to the step and diffuse along its upper edge without ever coming back to the cluster's interior. A potential energy profile which might result in such a diffusion pattern is sketched in Fig.~\ref{fig:1}b (e.g. see \cite{Golzhauser1996}). The novelty of the observed behavior has provoked a series of further studies on the subject \cite{Fu1998,*Oh2003} and it has been suggested that the observed effects might be a consequence of unusual relaxation patterns. Furthermore, similar phenomena have been observed in other surface systems. E.g., steps on Pt surfaces were found to facilitate the dissociation of $\mathrm{O}_2$ \cite{Gambardella2001} and CO \cite{Vidal-Iglesias2009} molecules and even promote water adsorption in step regions \cite{Morgenstern1996}. Yet a definitive answer as to the nature of such curious behavior has not been given.

    Here we demonstrate, that (i) the favorable edge adsorption can only be fully accounted for by taking into consideration the relativity of \emph{5d} elements (ii) but the width of the step-edge repulsion barrier on (111) surfaces of \emph{5d} metals is indeed dependent on atomic relaxations in the system. As a model system we consider Pt on Pt(111) and generalize the obtained results to include miscellaneous other \emph{5d} and \emph{s-p} species.

\begin{figure}
		\includegraphics{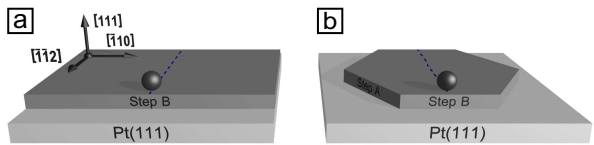}
        \caption{\label{fig:2} (color online) System setup for VASP (a) and KKR (b) calculations. (a) A stripe 14 atomic rows wide aligned along the $\sbr{\overline{1}10}$ direction of a Pt(111) surface. (b) Hexagonal Pt islands of $\sim 25$~\AA~ in diameter adsorbed homoepitaxially on a clean Pt(111) surface, a single atom adsorbed along the shorter diameter of the island ($\sbr{\overline{1}\overline{1}2}$ direction).}
	\end{figure}

    To check whether atomic relaxations (as proposed in \cite{Golzhauser1996,*Wang1991,*Wang1993}) could indeed account for the behavior described above and at the same time have a flexible approach to relativity we have made use of two different \emph{ab initio} approaches. To trace the effect of relaxations we have relied on the Vienna \emph{ab initio} simulation package (VASP) \cite{Kresse1993,*Kresse1996} where for geometry optimization a criterion of force-on-nuclei convergence to within 0.01~eV/\AA~was used. We have treated exchange and correlation effects with the local density approximation \cite{Ceperley1980} (LDA) and have used projector augmented wave (PAW) potentials \cite{Blochl1994}. The details of the calculational routine can be found in Refs. \cite{StepanyukO2009,Tian2010}. To model the experimental geometry the following setup has been used in our calculations. As the simulation of real-size islands in VASP is hardly possible from the computational point of view we have approximated experimentally observed Pt islands \cite{Golzhauser1996,*Wang1991,*Wang1993,Fu1998,*Oh2003} by an infinite stripe 14 atomic rows wide aligned along the $\sbr{\overline{1}10}$ direction of a Pt(111) surface (Fig.~\ref{fig:2}a). Single adatoms have been adsorbed atop the stripe along the $\sbr{\overline{1}\overline{1}2}$ line spanning the stripe from one step (type A) to the other (type B). Both fully relaxed and unrelaxed geometries have been considered.

    To obtain a better control over relativistic corrections we have employed an \emph{ab initio} Korringa-Kohn-Rostoker (KKR) Green's function method in the atomic spheres approximation  \cite{Wildberger1995,*Zeller1995}. This method gives a proper description of physics on surfaces with atomic steps \cite{Ignatiev2007} and islands \cite{Smirnov2008}. Normally the calculations are carried out under the inclusion of scalar-relativistic corrections to the Kohn-Sham equations. By excluding these corrections and carrying out fully self consistent calculations we were able to compare our system to its non-relativistic counterpart, thus underlining the governing role of relativity in the effects we describe. For KKR calculations we have constructed hexagonal Pt islands of $\sim 25$~\AA\ in diameter adsorbed homoepitaxially on a clean Pt(111) surface (Fig.~\ref{fig:2}b). To study adsorption energies, single atoms have then been adsorbed along the shorter diameter of the island ($\sbr{\overline{1}\overline{1}2}$ direction).


    \begin{figure}
		\includegraphics{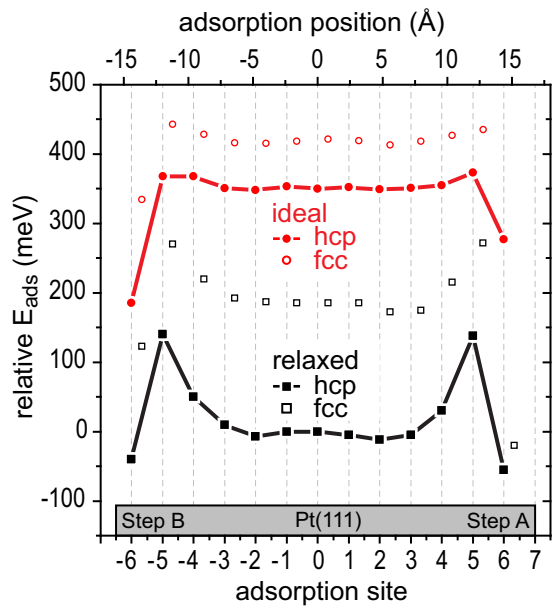}
        \caption{\label{fig:3} (color online) Adsorption energies of a Pt adatom diffusing in a straight line across a Pt(111) stripe for a fully relaxed (black squares) and ideal (red circles) geometries. Atom position is given in numbered adsorption sites and \AA~with respect to the center of the stripe. Integer numbers (filled squares and circles) denote \emph{fcc} and non-integer numbers (empty squares and circles) correspond to \emph{hcp} sites.}
	\end{figure}

    We start our investigation of the unusual energetics found in experiments \cite{Golzhauser1996,*Wang1991,*Wang1993,Fu1998,*Oh2003} by attempting to reproduce experimental observations in our VASP calculations. Let us first consider the adsorption of a Pt adatom diffusing across a Pt stripe (Fig.~\ref{fig:2}a). Corresponding energies are presented in Fig.~\ref{fig:3} for a fully relaxed (black squares) and ideal (red circles) geometries. As the energy is determined up to an additive constant we present the energies respective to the adsorption energy in the center of the stripe in the relaxed configuration. The unrelaxed values have been further offset by 350~meV for clarity. Upon studying both relaxed and unrelaxed configurations it becomes evident, that there is, indeed, an energetically unfavorable zone between the center of the cluster and its exterior. For the relaxed configuration the height of the barrier an adatom has to overcome to migrate from the center to the edge is of the order of 150~meV. Furthermore, in complete accord with the experiment \cite{Golzhauser1996,*Wang1991,*Wang1993,Fu1998,*Oh2003}, adsorption at the edge of the island is more favorable by about 50~meV than that at the center. Comparing the energy profiles for ideal and relaxed geometries one notices that the presence of structural relaxations strengthens the center-edge barrier effectively increasing its width. The results of KKR calculation further support the above-mentioned trends by yielding an adsorption energy at the B-step edge that is by 120~meV more favorable than the adsorption energy at the island's interior (the respective value for the A-step is 90~meV). If we consider the obtained energy differences in terms of Boltzmann statistics, adsorption at the step edge (at experimental temperatures of 100 K \cite{Golzhauser1996}) is by a factor of 300 more probable, than that at the center of a stripe or a cluster \footnote{$p_2/p_1=exp[-(E_2-E_1)/kT]$, where $p_{i}$ and $E_{i}$ are the probability to find an adsorbed atom at site $i$ and the adsorption energy at the corresponding site.}.

    By studying the energy profiles in Fig.~\ref{fig:3} we come to an evident conclusion that while atomic relaxations can play a role in defining the width of the edge repulsion barrier, which supports the hypothesis of Ehrlich and coworkers \cite{Golzhauser1996,*Wang1991,*Wang1993}, they can in no way account for the favorable adsorption at the step edge.

    Having ruled out the possibility of relaxations being the cause of preferable edge adsorption we have to ask ourselves is what else distinguishes \emph{5d} metal steps from steps of other metallic systems. Indeed the remaining most obvious difference is the high mass of \emph{5d} elements, which makes relativistic effects most pronounced, and is known to cause interesting surface related effects \cite{[{See, e.g. }][{}]Fiorentini1993,*Filippetti1999,*Mura2003}.

    The importance of relativistic effects for heavy-element chemistry and physics has been extensively discussed by Pyykk\"o and Desclaux \cite{Pyykko1979,[{}][{ and numerous references therein.}]Pyykko1988}. In terms of electronic structure relativi breaks down into several major effects \cite{Pyykko1979,*Pyykko1988,Christensen1984}: (i) Electrons of \emph{s} (and partially \emph{p}) character experience the relativistic mass effect which causes the respective shells to contract \cite{Pyykko1979}. For outer \emph{s} and \emph{p} shells the effect is somewhat indirect as the relativistic contraction is only experienced by the inner part of the wavefunction and the outer tails are pulled in as a consequence; (ii) palpable core levels' contraction leads to a stronger screening of the core, which decreases the pull on the \emph{d} and \emph{f} atomic orbitals causing them to expand radially and destabilize energetically \cite{Pyykko1979}; (iii) the last effect is the spin-orbit coupling (SOC), which lifts the degeneracy of electrons with the same quantum number \emph{l}. The SOC-related energy splitting may reach up to several meV \cite{Gambardella2003,Pyykko1988}, yet this is still at least an order of magnitude smaller than the energy differences which we are trying to explain, so we will neglect the SOC for the rest of the present study. Effects (i) and (ii) can be accounted for by using scalar-relativistic corrections.

    \begin{figure}
		\includegraphics{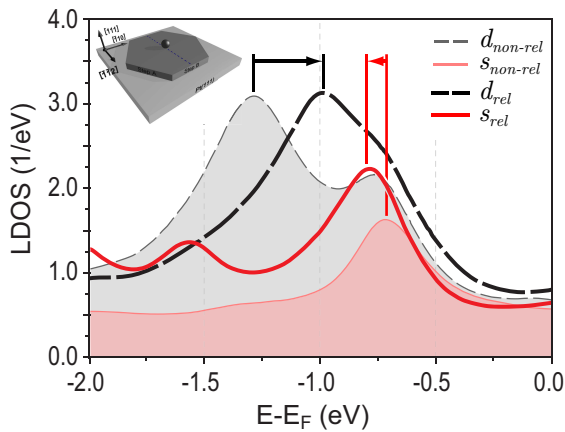}
        \caption{\label{fig:4} (color online) LDOS at a single Pt adatom residing at the center of a hexagonal Pt(111) island calculated non-relativistically (thin filled curves) and with scalar-relativistic corrections (thick solid curves) for \emph{s} (red solid curves) and \emph{d} (black broken curves) electrons. The \emph{s} electron densities are scaled up by a factor of 20 for clarity.}
	\end{figure}

    Let us now see whether we can apply our knowledge about relativistic effects to explain the favorable step-edge adsorption on Pt islands. First of all we have to check what changes (in terms of the electronic structure) relativity introduces into our system. Let us consider a single Pt atom adsorbed at the center of a hexagonal Pt(111) island some 25~\AA~in diameter. To separate the effects of relativity we perform KKR calculations both with scalar-relativistic corrections included and without them. The resulting local densities of \emph{s} and \emph{d} electron states (LDOS) for both cases are presented in Fig.~\ref{fig:4}. Thin filled curves represent the non-relativistic LDOS while thick solid curves reflect for the relativistic case (in both cases red solid curves denote the density of \emph{s} electrons scaled up by a factor of 20 for clarity, and black broken curves represent the \emph{d}-LDOS). Straight comparison immediately reveals both the \emph{s}-level contraction (by about 80~meV for the level localized around $-0.75$~eV) and the self-consistent \emph{d}-level expansion (by about 300~meV for the level localized at $-1.0$~eV in the relativistic case). In accord with our expectations, the effect of relativity on the outer \emph{s}-shell is weaker than the indirect effect of the core screening on the \emph{d}-level.

    Such energy shifts of \emph{s} and \emph{d} states bring them closer together thus increasing their overlap and consequently the \emph{s-d} hybridization. Much stronger, relativistically enhanced, \emph{s-d} bonding can well account for the difference in adsorption behavior between the elements of \emph{5d} and \emph{3d}-\emph{4d} ranges. Indeed it is known, that the strength of the \emph{s-d} hybridization can well decide over the preferences of bonding geometry of both relativistic and non-relativistic elements \cite{Hakkinen2002}.

\begin{figure}
		\includegraphics{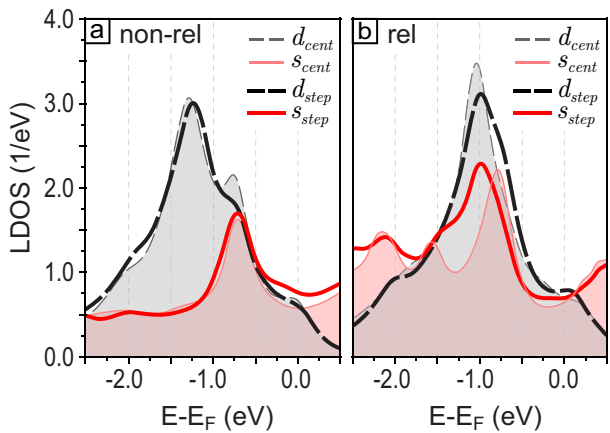}
        \caption{\label{fig:5} (color online) (a) LDOS at a single Pt adatom residing at the center (thin filled curves, red solid for \emph{s} and black broken for \emph{d} electrons) and at the B-step edge (thick curves, red solid for \emph{s} and black broken for \emph{d} electrons) of a hexagonal Pt(111) island, calculated non-relativistically. (b) The same LDOS calculated with scalar-relativistic corrections. The \emph{s} electron densities are scaled up by a factor of 20 for clarity.}
	\end{figure}

    To understand the impact of relativity and check our hypothesis about the predominant role of the \emph{s-d} hybridization in the adsorption of relativistic impurities on relativistic substrates, let us consider, how the LDOS and the adsorption energy evolve as we move our Pt atom adsorbed on a hexagonal Pt(111) island from the island's center to the edge (Fig.~\ref{fig:5}). If we neglect the scalar-relativistic corrections the \emph{s} and \emph{d} LDOS of a Pt adatom at the center of the Pt island (red solid and black broken thin filled curves in Fig.~\ref{fig:5}a respectively) remains almost unchanged as we move the atom to a B-step edge of the island (thick solid curves in Fig.~\ref{fig:5}, red solid for \emph{s}-LDOS and black broken for \emph{d}-LDOS). The only tangible effect is a slight destabilization of the \emph{d} level and a slight reduction of the \emph{s-d} overlap, both of which are expectable due to the reduced coordination \cite{Mavropoulos2003,Hakkinen2002}. As a result, adsorption at the B-step for non-relativistic Pt is 80~meV less favorable than at the center. If we, however, include the relativistic corrections in our calculations the result is strikingly different. Already at the center of the island the \emph{s} and \emph{d} electronic levels of the Pt adsorbate are pushed closer together by the relativistic effects thus increasing the \emph{s-d} overlap. But as we move the atom towards the step we find the both \emph{s} and \emph{d} levels are yet again shifted which lands them at the same energy and effectively locks them together. As a result, the \emph{s-d} hybridization is maximized reducing the adsorption energy of the adatom and making edge adsorption favorable.

    Now only one questions remains to be resolved: why do the electronic levels shift upon moving the adatom to the edge in the relativistic case and remain almost unchanged when the relativity is shut down. This difference can be best understood if one considers the bond formation at the center of the island and at its edge self-consistently. Reduced coordination at the edge reduces the number of bonds weakening the overall binding of the adsorbate to the substrate while making each of the remaining bonds a bit stronger as one would expect in the context of Pauling's rules \cite{Pauling1929}. In the non-relativistic case the outer \emph{d} orbitals are stable and cannot profit from the slight increase of single bonds' strengths and thus can not counteract the weakening of the overall bonding strength and the reduction of the \emph{s} and \emph{p} electron density due to the Smoluchowski effect. In the relativistic case, however, the \emph{d} orbitals are destabilized by the core screening and thus can more actively participate in the bonding (both \emph{s-d} and \emph{d-d}). In this way the adatom is able to profit from the bond enhancement by increasing the \emph{s-d} overlap with the neighboring atom's shells, which in turn, self-consistently, brings the \emph{s} and \emph{d} levels closer together until they ``lock'' at the same energy, producing the LDOS displayed in Fig.~\ref{fig:5}b.

    The feasibility of such an adsorption and bonding scenario is further supported by the fact, that mixed systems, e.g. a \emph{3d} adatom on a \emph{5d} substrate, show an intermediate behavior, i.e. the \emph{s} and \emph{d} orbitals are more susceptible to the adsorption geometry changes than in non-relativistic systems, yet the introduced shifts of \emph{s} and \emph{d} levels are smaller than e.g. in Pt/Pt(111). For example, for Cu/Pt(111) the \emph{s}-\emph{d} interlevel energy difference is reduced by 25\% upon the shift of the adatom from the center of the island to its edge. The \emph{s-d} overlap is thus slightly increased. However, this increase turns out to be insufficient to afford edge adsorption: according to our calculations edge adsorption for Cu/Pt(111) is by about 100 meV less favorable than the adsorption on a clean surface or in the center of an island.

    If we stray from our model system of Pt on Pt(111) we would also notice, that similar behavior can be found in many systems where relativity is present or the outer orbitals of adsorbates are unstable enough to be strongly affected by the bonding strength increase at a step edge. For example, our VASP calculations show that while \emph{3d} and \emph{4d} elements (Cu, Au and Pd) tend to adsorb more readily in the interior of a Pt(111) island, relativistic \emph{5d} species (Pt, Ir) and species of a more pronounced \emph{s-p} character (C, O, N) follow suit of Pt and favor step edge adsorption.

    We might thus conclude that the curious diffusion patterns observed in experiments in Refs.~\cite{Golzhauser1996,*Wang1991,*Wang1993,Fu1998,*Oh2003} is a consequence of a delicate balance where on one side of the scales the Smoluchowski effect and coordination reduction strive to weaken the bonding of adsorbates to the substrate and on the other side the individual bonds' enhancement striving to outweigh them. Now we have shown that while in non-relativistic system the edge adsorption is unfavorable, relativistic effects are well able to topple the balance.

    Apart of the fundamental value of this notion, we believe that it might prove to be applicable in surface engineering, e.g. for the bottom-up creation of various surface based nanostructures for further electronic and nanocatalytic research \cite{Gambardella2001,Vidal-Iglesias2009}.


\begin{thebibliography}{36}%
    \makeatletter
    \providecommand \@ifxundefined [1]{%
     \@ifx{#1\undefined}
    }%
    \providecommand \@ifnum [1]{%
     \ifnum #1\expandafter \@firstoftwo
     \else \expandafter \@secondoftwo
     \fi
    }%
    \providecommand \@ifx [1]{%
     \ifx #1\expandafter \@firstoftwo
     \else \expandafter \@secondoftwo
     \fi
    }%
    \providecommand \natexlab [1]{#1}%
    \providecommand \enquote  [1]{``#1''}%
    \providecommand \bibnamefont  [1]{#1}%
    \providecommand \bibfnamefont [1]{#1}%
    \providecommand \citenamefont [1]{#1}%
    \providecommand \href@noop [0]{\@secondoftwo}%
    \providecommand \href [0]{\begingroup \@sanitize@url \@href}%
    \providecommand \@href[1]{\@@startlink{#1}\@@href}%
    \providecommand \@@href[1]{\endgroup#1\@@endlink}%
    \providecommand \@sanitize@url [0]{\catcode `\\12\catcode `\$12\catcode
      `\&12\catcode `\#12\catcode `\^12\catcode `\_12\catcode `\%12\relax}%
    \providecommand \@@startlink[1]{}%
    \providecommand \@@endlink[0]{}%
    \providecommand \url  [0]{\begingroup\@sanitize@url \@url }%
    \providecommand \@url [1]{\endgroup\@href {#1}{\urlprefix }}%
    \providecommand \urlprefix  [0]{URL }%
    \providecommand \Eprint [0]{\href }%
    \@ifxundefined \urlstyle {%
      \providecommand \doi  [0]{\begingroup \@sanitize@url \@doi}%
      \providecommand \@doi [1]{\endgroup \@@startlink {\doibase
      #1}doi:\discretionary {}{}{}#1\@@endlink }%
    }{%
      \providecommand \doi  [0]{doi:\discretionary{}{}{}\begingroup
      \urlstyle{rm}\Url }%
    }%
    \providecommand \doibase [0]{http://dx.doi.org/}%
    \providecommand \Doi [0]{\begingroup \@sanitize@url \@Doi }%
    \providecommand \@Doi  [1]{\endgroup\@@startlink{\doibase#1}\@@Doi}%
    \providecommand \@@Doi [1]{#1\@@endlink}%
    \providecommand \selectlanguage [0]{\@gobble}%
    \providecommand \bibinfo  [0]{\@secondoftwo}%
    \providecommand \bibfield  [0]{\@secondoftwo}%
    \providecommand \translation [1]{[#1]}%
    \providecommand \BibitemOpen [0]{}%
    \providecommand \bibitemStop [0]{}%
    \providecommand \bibitemNoStop [0]{.\EOS\space}%
    \providecommand \EOS [0]{\spacefactor3000\relax}%
    \providecommand \BibitemShut  [1]{\csname bibitem#1\endcsname}%
    \bibitem [{\citenamefont {G\"olzh\"auser}\ and\ \citenamefont
      {Ehrlich}(1996)}]{Golzhauser1996}%
      \BibitemOpen
      \bibfield  {author} {\bibinfo {author} {\bibfnamefont {A.}~\bibnamefont
      {G\"olzh\"auser}}\ and\ \bibinfo {author} {\bibfnamefont {G.}~\bibnamefont
      {Ehrlich}},\ }\Doi {10.1103/PhysRevLett.77.1334} {\bibfield  {journal}
      {\bibinfo  {journal} {Phys. Rev. Lett.},\ }\textbf {\bibinfo {volume} {77}},\
      \bibinfo {pages} {1334} (\bibinfo {year} {1996})}\BibitemShut {NoStop}%
    \bibitem [{\citenamefont {Wang}\ and\ \citenamefont
      {Ehrlich}(1991)}]{Wang1991}%
      \BibitemOpen
      \bibfield  {author} {\bibinfo {author} {\bibfnamefont {S.~C.}\ \bibnamefont
      {Wang}}\ and\ \bibinfo {author} {\bibfnamefont {G.}~\bibnamefont {Ehrlich}},\
      }\Doi {10.1103/PhysRevLett.67.2509} {\bibfield  {journal} {\bibinfo
      {journal} {Phys. Rev. Lett.},\ }\textbf {\bibinfo {volume} {67}},\ \bibinfo
      {pages} {2509} (\bibinfo {year} {1991})}\BibitemShut {NoStop}%
    \bibitem [{\citenamefont {Wang}\ and\ \citenamefont
      {Ehrlich}(1993)}]{Wang1993}%
      \BibitemOpen
      \bibfield  {author} {\bibinfo {author} {\bibfnamefont {S.~C.}\ \bibnamefont
      {Wang}}\ and\ \bibinfo {author} {\bibfnamefont {G.}~\bibnamefont {Ehrlich}},\
      }\Doi {10.1103/PhysRevLett.70.41} {\bibfield  {journal} {\bibinfo  {journal}
      {Phys. Rev. Lett.},\ }\textbf {\bibinfo {volume} {70}},\ \bibinfo {pages}
      {41} (\bibinfo {year} {1993})}\BibitemShut {NoStop}%
    \bibitem [{\citenamefont {Ehrlich}\ and\ \citenamefont
      {Hudda}(1966)}]{Ehrlich1966}%
      \BibitemOpen
      \bibfield  {author} {\bibinfo {author} {\bibfnamefont {G.}~\bibnamefont
      {Ehrlich}}\ and\ \bibinfo {author} {\bibfnamefont {F.~G.}\ \bibnamefont
      {Hudda}},\ }\Doi {10.1063/1.1726787} {\bibfield  {journal} {\bibinfo
      {journal} {The Journal of Chemical Physics},\ }\textbf {\bibinfo {volume}
      {44}},\ \bibinfo {pages} {1039} (\bibinfo {year} {1966})}\BibitemShut
      {NoStop}%
    \bibitem [{\citenamefont {Schwoebel}\ and\ \citenamefont
      {Shipsey}(1966)}]{Schwoebel1966}%
      \BibitemOpen
      \bibfield  {author} {\bibinfo {author} {\bibfnamefont {R.~L.}\ \bibnamefont
      {Schwoebel}}\ and\ \bibinfo {author} {\bibfnamefont {E.~J.}\ \bibnamefont
      {Shipsey}},\ }\Doi {10.1063/1.1707904} {\bibfield  {journal} {\bibinfo
      {journal} {Journal of Applied Physics},\ }\textbf {\bibinfo {volume} {37}},\
      \bibinfo {pages} {3682} (\bibinfo {year} {1966})}\BibitemShut {NoStop}%
    \bibitem [{\citenamefont {Mo}\ \emph {et~al.}(2008)\citenamefont {Mo},
      \citenamefont {Zhu}, \citenamefont {Kaxiras},\ and\ \citenamefont
      {Zhang}}]{Mo2008}%
      \BibitemOpen
      \bibfield  {author} {\bibinfo {author} {\bibfnamefont {Y.}~\bibnamefont
      {Mo}}, \bibinfo {author} {\bibfnamefont {W.}~\bibnamefont {Zhu}}, \bibinfo
      {author} {\bibfnamefont {E.}~\bibnamefont {Kaxiras}}, \ and\ \bibinfo
      {author} {\bibfnamefont {Z.}~\bibnamefont {Zhang}},\ }\Doi
      {10.1103/PhysRevLett.101.216101} {\bibfield  {journal} {\bibinfo  {journal}
      {Phys. Rev. Lett.},\ }\textbf {\bibinfo {volume} {101}},\ \bibinfo {pages}
      {216101} (\bibinfo {year} {2008})}\BibitemShut {NoStop}%
    \bibitem [{\citenamefont {Smoluchowski}(1941)}]{Smoluchowski1941}%
      \BibitemOpen
      \bibfield  {author} {\bibinfo {author} {\bibfnamefont {R.}~\bibnamefont
      {Smoluchowski}},\ }\Doi {10.1103/PhysRev.60.661} {\bibfield  {journal}
      {\bibinfo  {journal} {Phys. Rev.},\ }\textbf {\bibinfo {volume} {60}},\
      \bibinfo {pages} {661} (\bibinfo {year} {1941})}\BibitemShut {NoStop}%
    \bibitem [{\citenamefont {Bromann}\ \emph {et~al.}(1995)\citenamefont
      {Bromann}, \citenamefont {Brune}, \citenamefont {R\"oder},\ and\
      \citenamefont {Kern}}]{Bromann1995}%
      \BibitemOpen
      \bibfield  {author} {\bibinfo {author} {\bibfnamefont {K.}~\bibnamefont
      {Bromann}}, \bibinfo {author} {\bibfnamefont {H.}~\bibnamefont {Brune}},
      \bibinfo {author} {\bibfnamefont {H.}~\bibnamefont {R\"oder}}, \ and\
      \bibinfo {author} {\bibfnamefont {K.}~\bibnamefont {Kern}},\ }\Doi
      {10.1103/PhysRevLett.75.677} {\bibfield  {journal} {\bibinfo  {journal}
      {Phys. Rev. Lett.},\ }\textbf {\bibinfo {volume} {75}},\ \bibinfo {pages}
      {677} (\bibinfo {year} {1995})}\BibitemShut {NoStop}%
    \bibitem [{\citenamefont {Morgenstern}\ \emph {et~al.}(1998)\citenamefont
      {Morgenstern}, \citenamefont {Rosenfeld}, \citenamefont {L\ae{}gsgaard},
      \citenamefont {Besenbacher},\ and\ \citenamefont {Comsa}}]{Morgenstern1998}%
      \BibitemOpen
      \bibfield  {author} {\bibinfo {author} {\bibfnamefont {K.}~\bibnamefont
      {Morgenstern}}, \bibinfo {author} {\bibfnamefont {G.}~\bibnamefont
      {Rosenfeld}}, \bibinfo {author} {\bibfnamefont {E.}~\bibnamefont
      {L\ae{}gsgaard}}, \bibinfo {author} {\bibfnamefont {F.}~\bibnamefont
      {Besenbacher}}, \ and\ \bibinfo {author} {\bibfnamefont {G.}~\bibnamefont
      {Comsa}},\ }\Doi {10.1103/PhysRevLett.80.556} {\bibfield  {journal} {\bibinfo
       {journal} {Phys. Rev. Lett.},\ }\textbf {\bibinfo {volume} {80}},\ \bibinfo
      {pages} {556} (\bibinfo {year} {1998})}\BibitemShut {NoStop}%
    \bibitem [{\citenamefont {Prieto}\ \emph {et~al.}(2000)\citenamefont {Prieto},
      \citenamefont {de~la Figuera},\ and\ \citenamefont {Miranda}}]{Prieto2000}%
      \BibitemOpen
      \bibfield  {author} {\bibinfo {author} {\bibfnamefont {J.~E.}\ \bibnamefont
      {Prieto}}, \bibinfo {author} {\bibfnamefont {J.}~\bibnamefont {de~la
      Figuera}}, \ and\ \bibinfo {author} {\bibfnamefont {R.}~\bibnamefont
      {Miranda}},\ }\Doi {10.1103/PhysRevB.62.2126} {\bibfield  {journal} {\bibinfo
       {journal} {Phys. Rev. B},\ }\textbf {\bibinfo {volume} {62}},\ \bibinfo
      {pages} {2126} (\bibinfo {year} {2000})}\BibitemShut {NoStop}%
    \bibitem [{\citenamefont {Fu}\ \emph {et~al.}(1998)\citenamefont {Fu},
      \citenamefont {Wu},\ and\ \citenamefont {Tsong}}]{Fu1998}%
      \BibitemOpen
      \bibfield  {author} {\bibinfo {author} {\bibfnamefont {T.-Y.}\ \bibnamefont
      {Fu}}, \bibinfo {author} {\bibfnamefont {H.-T.}\ \bibnamefont {Wu}}, \ and\
      \bibinfo {author} {\bibfnamefont {T.~T.}\ \bibnamefont {Tsong}},\ }\Doi
      {10.1103/PhysRevB.58.2340} {\bibfield  {journal} {\bibinfo  {journal} {Phys.
      Rev. B},\ }\textbf {\bibinfo {volume} {58}},\ \bibinfo {pages} {2340}
      (\bibinfo {year} {1998})}\BibitemShut {NoStop}%
    \bibitem [{\citenamefont {Oh}\ \emph {et~al.}(2003)\citenamefont {Oh},
      \citenamefont {Kyuno}, \citenamefont {Wang},\ and\ \citenamefont
      {Ehrlich}}]{Oh2003}%
      \BibitemOpen
      \bibfield  {author} {\bibinfo {author} {\bibfnamefont {S.-M.}\ \bibnamefont
      {Oh}}, \bibinfo {author} {\bibfnamefont {K.}~\bibnamefont {Kyuno}}, \bibinfo
      {author} {\bibfnamefont {S.~C.}\ \bibnamefont {Wang}}, \ and\ \bibinfo
      {author} {\bibfnamefont {G.}~\bibnamefont {Ehrlich}},\ }\Doi
      {10.1103/PhysRevB.67.075413} {\bibfield  {journal} {\bibinfo  {journal}
      {Phys. Rev. B},\ }\textbf {\bibinfo {volume} {67}},\ \bibinfo {pages}
      {075413} (\bibinfo {year} {2003})}\BibitemShut {NoStop}%
    \bibitem [{\citenamefont {Gambardella}\ \emph {et~al.}(2001)\citenamefont
      {Gambardella}, \citenamefont {\ifmmode \check{S}\else
      \v{S}\fi{}ljivan\ifmmode~\check{c}\else \v{c}\fi{}anin}, \citenamefont
      {Hammer}, \citenamefont {Blanc}, \citenamefont {Kuhnke},\ and\ \citenamefont
      {Kern}}]{Gambardella2001}%
      \BibitemOpen
      \bibfield  {author} {\bibinfo {author} {\bibfnamefont {P.}~\bibnamefont
      {Gambardella}}, \bibinfo {author} {\bibfnamefont {{\ifmmode \check{Z}\else
      \v{Z}\fi}.}~\bibnamefont {\ifmmode \check{S}\else
      \v{S}\fi{}ljivan\ifmmode~\check{c}\else \v{c}\fi{}anin}}, \bibinfo {author}
      {\bibfnamefont {B.}~\bibnamefont {Hammer}}, \bibinfo {author} {\bibfnamefont
      {M.}~\bibnamefont {Blanc}}, \bibinfo {author} {\bibfnamefont
      {K.}~\bibnamefont {Kuhnke}}, \ and\ \bibinfo {author} {\bibfnamefont
      {K.}~\bibnamefont {Kern}},\ }\Doi {10.1103/PhysRevLett.87.056103} {\bibfield
      {journal} {\bibinfo  {journal} {Phys. Rev. Lett.},\ }\textbf {\bibinfo
      {volume} {87}},\ \bibinfo {pages} {056103} (\bibinfo {year}
      {2001})}\BibitemShut {NoStop}%
    \bibitem [{\citenamefont {Vidal-Iglesias}\ \emph {et~al.}(2009)\citenamefont
      {Vidal-Iglesias}, \citenamefont {Solla-Gull\`on}, \citenamefont
      {{Campi\~{n}a}}, \citenamefont {Herrero}, \citenamefont {Aldaz},\ and\
      \citenamefont {Feliu}}]{Vidal-Iglesias2009}%
      \BibitemOpen
      \bibfield  {author} {\bibinfo {author} {\bibfnamefont {F.}~\bibnamefont
      {Vidal-Iglesias}}, \bibinfo {author} {\bibfnamefont {J.}~\bibnamefont
      {Solla-Gull\`on}}, \bibinfo {author} {\bibfnamefont {J.}~\bibnamefont
      {{Campi\~{n}a}}}, \bibinfo {author} {\bibfnamefont {E.}~\bibnamefont
      {Herrero}}, \bibinfo {author} {\bibfnamefont {A.}~\bibnamefont {Aldaz}}, \
      and\ \bibinfo {author} {\bibfnamefont {J.}~\bibnamefont {Feliu}},\ }\Doi
      {DOI: 10.1016/j.electacta.2009.03.025} {\bibfield  {journal} {\bibinfo
      {journal} {Electrochim. Acta},\ }\textbf {\bibinfo {volume} {54}},\ \bibinfo
      {pages} {4459 } (\bibinfo {year} {2009})}\BibitemShut {NoStop}%
    \bibitem [{\citenamefont {Morgenstern}\ \emph {et~al.}(1996)\citenamefont
      {Morgenstern}, \citenamefont {Michely},\ and\ \citenamefont
      {Comsa}}]{Morgenstern1996}%
      \BibitemOpen
      \bibfield  {author} {\bibinfo {author} {\bibfnamefont {M.}~\bibnamefont
      {Morgenstern}}, \bibinfo {author} {\bibfnamefont {T.}~\bibnamefont
      {Michely}}, \ and\ \bibinfo {author} {\bibfnamefont {G.}~\bibnamefont
      {Comsa}},\ }\Doi {10.1103/PhysRevLett.77.703} {\bibfield  {journal} {\bibinfo
       {journal} {Phys. Rev. Lett.},\ }\textbf {\bibinfo {volume} {77}},\ \bibinfo
      {pages} {703} (\bibinfo {year} {1996})}\BibitemShut {NoStop}%
    \bibitem [{\citenamefont {Kresse}\ and\ \citenamefont
      {Hafner}(1993)}]{Kresse1993}%
      \BibitemOpen
      \bibfield  {author} {\bibinfo {author} {\bibfnamefont {G.}~\bibnamefont
      {Kresse}}\ and\ \bibinfo {author} {\bibfnamefont {J.}~\bibnamefont
      {Hafner}},\ }\Doi {10.1103/PhysRevB.47.558} {\bibfield  {journal} {\bibinfo
      {journal} {Phys. Rev. B},\ }\textbf {\bibinfo {volume} {47}},\ \bibinfo
      {pages} {558} (\bibinfo {year} {1993})}\BibitemShut {NoStop}%
    \bibitem [{\citenamefont {Kresse}\ and\ \citenamefont
      {Furthm\"uller}(1996)}]{Kresse1996}%
      \BibitemOpen
      \bibfield  {author} {\bibinfo {author} {\bibfnamefont {G.}~\bibnamefont
      {Kresse}}\ and\ \bibinfo {author} {\bibfnamefont {J.}~\bibnamefont
      {Furthm\"uller}},\ }\Doi {10.1103/PhysRevB.54.11169} {\bibfield  {journal}
      {\bibinfo  {journal} {Phys. Rev. B},\ }\textbf {\bibinfo {volume} {54}},\
      \bibinfo {pages} {11169} (\bibinfo {year} {1996})}\BibitemShut {NoStop}%
    \bibitem [{\citenamefont {Ceperley}\ and\ \citenamefont
      {Alder}(1980)}]{Ceperley1980}%
      \BibitemOpen
      \bibfield  {author} {\bibinfo {author} {\bibfnamefont {D.~M.}\ \bibnamefont
      {Ceperley}}\ and\ \bibinfo {author} {\bibfnamefont {B.~J.}\ \bibnamefont
      {Alder}},\ }\Doi {10.1103/PhysRevLett.45.566} {\bibfield  {journal} {\bibinfo
       {journal} {Phys. Rev. Lett.},\ }\textbf {\bibinfo {volume} {45}},\ \bibinfo
      {pages} {566} (\bibinfo {year} {1980})}\BibitemShut {NoStop}%
    \bibitem [{\citenamefont {Bl\"ochl}(1994)}]{Blochl1994}%
      \BibitemOpen
      \bibfield  {author} {\bibinfo {author} {\bibfnamefont {P.~E.}\ \bibnamefont
      {Bl\"ochl}},\ }\Doi {10.1103/PhysRevB.50.17953} {\bibfield  {journal}
      {\bibinfo  {journal} {Phys. Rev. B},\ }\textbf {\bibinfo {volume} {50}},\
      \bibinfo {pages} {17953} (\bibinfo {year} {1994})}\BibitemShut {NoStop}%
    \bibitem [{\citenamefont {Stepanyuk}\ \emph {et~al.}(2009)\citenamefont
      {Stepanyuk}, \citenamefont {Negulyaev}, \citenamefont {Ignatiev},
      \citenamefont {Przybylski}, \citenamefont {Hergert}, \citenamefont
      {Saletsky},\ and\ \citenamefont {Kirschner}}]{StepanyukO2009}%
      \BibitemOpen
      \bibfield  {author} {\bibinfo {author} {\bibfnamefont {O.~V.}\ \bibnamefont
      {Stepanyuk}}, \bibinfo {author} {\bibfnamefont {N.~N.}\ \bibnamefont
      {Negulyaev}}, \bibinfo {author} {\bibfnamefont {P.~A.}\ \bibnamefont
      {Ignatiev}}, \bibinfo {author} {\bibfnamefont {M.}~\bibnamefont
      {Przybylski}}, \bibinfo {author} {\bibfnamefont {W.}~\bibnamefont {Hergert}},
      \bibinfo {author} {\bibfnamefont {A.~M.}\ \bibnamefont {Saletsky}}, \ and\
      \bibinfo {author} {\bibfnamefont {J.}~\bibnamefont {Kirschner}},\ }\Doi
      {10.1103/PhysRevB.79.155410} {\bibfield  {journal} {\bibinfo  {journal}
      {Phys. Rev. B},\ }\textbf {\bibinfo {volume} {79}},\ \bibinfo {pages}
      {155410} (\bibinfo {year} {2009})}\BibitemShut {NoStop}%
    \bibitem [{\citenamefont {Tian}\ \emph {et~al.}(2010)\citenamefont {Tian},
      \citenamefont {Sander}, \citenamefont {Negulyaev}, \citenamefont
      {Stepanyuk},\ and\ \citenamefont {Kirschner}}]{Tian2010}%
      \BibitemOpen
      \bibfield  {author} {\bibinfo {author} {\bibfnamefont {Z.}~\bibnamefont
      {Tian}}, \bibinfo {author} {\bibfnamefont {D.}~\bibnamefont {Sander}},
      \bibinfo {author} {\bibfnamefont {N.~N.}\ \bibnamefont {Negulyaev}}, \bibinfo
      {author} {\bibfnamefont {V.~S.}\ \bibnamefont {Stepanyuk}}, \ and\ \bibinfo
      {author} {\bibfnamefont {J.}~\bibnamefont {Kirschner}},\ }\Doi
      {10.1103/PhysRevB.81.113407} {\bibfield  {journal} {\bibinfo  {journal}
      {Phys. Rev. B},\ }\textbf {\bibinfo {volume} {81}},\ \bibinfo {pages}
      {113407} (\bibinfo {year} {2010})}\BibitemShut {NoStop}%
    \bibitem [{\citenamefont {Wildberger}\ \emph {et~al.}(1995)\citenamefont
      {Wildberger}, \citenamefont {Stepanyuk}, \citenamefont {Lang}, \citenamefont
      {Zeller},\ and\ \citenamefont {Dederichs}}]{Wildberger1995}%
      \BibitemOpen
      \bibfield  {author} {\bibinfo {author} {\bibfnamefont {K.}~\bibnamefont
      {Wildberger}}, \bibinfo {author} {\bibfnamefont {V.~S.}\ \bibnamefont
      {Stepanyuk}}, \bibinfo {author} {\bibfnamefont {P.}~\bibnamefont {Lang}},
      \bibinfo {author} {\bibfnamefont {R.}~\bibnamefont {Zeller}}, \ and\ \bibinfo
      {author} {\bibfnamefont {P.~H.}\ \bibnamefont {Dederichs}},\ }\Doi
      {10.1103/PhysRevLett.75.509} {\bibfield  {journal} {\bibinfo  {journal}
      {Phys. Rev. Lett.},\ }\textbf {\bibinfo {volume} {75}},\ \bibinfo {pages}
      {509} (\bibinfo {year} {1995})}\BibitemShut {NoStop}%
    \bibitem [{\citenamefont {Zeller}\ \emph {et~al.}(1995)\citenamefont {Zeller},
      \citenamefont {Dederichs}, \citenamefont {{\'U}jfalussy}, \citenamefont
      {Szunyogh},\ and\ \citenamefont {Weinberger}}]{Zeller1995}%
      \BibitemOpen
      \bibfield  {author} {\bibinfo {author} {\bibfnamefont {R.}~\bibnamefont
      {Zeller}}, \bibinfo {author} {\bibfnamefont {P.~H.}\ \bibnamefont
      {Dederichs}}, \bibinfo {author} {\bibfnamefont {B.}~\bibnamefont
      {{\'U}jfalussy}}, \bibinfo {author} {\bibfnamefont {L.}~\bibnamefont
      {Szunyogh}}, \ and\ \bibinfo {author} {\bibfnamefont {P.}~\bibnamefont
      {Weinberger}},\ }\Doi {10.1103/PhysRevB.52.8807} {\bibfield  {journal}
      {\bibinfo  {journal} {Phys. Rev. B},\ }\textbf {\bibinfo {volume} {52}},\
      \bibinfo {pages} {8807} (\bibinfo {year} {1995})}\BibitemShut {NoStop}%
    \bibitem [{\citenamefont {Ignatiev}\ \emph {et~al.}(2007)\citenamefont
      {Ignatiev}, \citenamefont {Stepanyuk}, \citenamefont {Klavsyuk},
      \citenamefont {Hergert},\ and\ \citenamefont {Bruno}}]{Ignatiev2007}%
      \BibitemOpen
      \bibfield  {author} {\bibinfo {author} {\bibfnamefont {P.~A.}\ \bibnamefont
      {Ignatiev}}, \bibinfo {author} {\bibfnamefont {V.~S.}\ \bibnamefont
      {Stepanyuk}}, \bibinfo {author} {\bibfnamefont {A.~L.}\ \bibnamefont
      {Klavsyuk}}, \bibinfo {author} {\bibfnamefont {W.}~\bibnamefont {Hergert}}, \
      and\ \bibinfo {author} {\bibfnamefont {P.}~\bibnamefont {Bruno}},\ }\Doi
      {10.1103/PhysRevB.75.155428} {\bibfield  {journal} {\bibinfo  {journal}
      {Phys. Rev. B},\ }\textbf {\bibinfo {volume} {75}},\ \bibinfo {pages}
      {155428} (\bibinfo {year} {2007})}\BibitemShut {NoStop}%
    \bibitem [{\citenamefont {Smirnov}\ \emph {et~al.}(2008)\citenamefont
      {Smirnov}, \citenamefont {Negulyaev}, \citenamefont {Niebergall},
      \citenamefont {Hergert}, \citenamefont {Saletsky},\ and\ \citenamefont
      {Stepanyuk}}]{Smirnov2008}%
      \BibitemOpen
      \bibfield  {author} {\bibinfo {author} {\bibfnamefont {A.~S.}\ \bibnamefont
      {Smirnov}}, \bibinfo {author} {\bibfnamefont {N.~N.}\ \bibnamefont
      {Negulyaev}}, \bibinfo {author} {\bibfnamefont {L.}~\bibnamefont
      {Niebergall}}, \bibinfo {author} {\bibfnamefont {W.}~\bibnamefont {Hergert}},
      \bibinfo {author} {\bibfnamefont {A.~M.}\ \bibnamefont {Saletsky}}, \ and\
      \bibinfo {author} {\bibfnamefont {V.~S.}\ \bibnamefont {Stepanyuk}},\ }\Doi
      {10.1103/PhysRevB.78.041405} {\bibfield  {journal} {\bibinfo  {journal}
      {Phys. Rev. B},\ }\textbf {\bibinfo {volume} {78}},\ \bibinfo {pages}
      {041405} (\bibinfo {year} {2008})}\BibitemShut {NoStop}%
    \bibitem [{Note1()}]{Note1}%
      \BibitemOpen
      \bibinfo {note} {$p_2/p_1=exp[-(E_2-E_1)/kT]$, where $p_{i}$ and $E_{i}$ are
      the probability to find an adsorbed atom at site $i$ and the adsorption
      energy at the corresponding site.}\BibitemShut {Stop}%
    \bibitem [{\citenamefont {Fiorentini}\ \emph {et~al.}(1993)\citenamefont
      {Fiorentini}, \citenamefont {Methfessel},\ and\ \citenamefont
      {Scheffler}}]{Fiorentini1993}%
      \BibitemOpen
      \bibfield  {author} {\bibinfo {author} {\bibfnamefont {V.}~\bibnamefont
      {Fiorentini}}, \bibinfo {author} {\bibfnamefont {M.}~\bibnamefont
      {Methfessel}}, \ and\ \bibinfo {author} {\bibfnamefont {M.}~\bibnamefont
      {Scheffler}},\ }\Doi {10.1103/PhysRevLett.71.1051} {\bibfield  {journal}
      {\bibinfo  {journal} {Phys. Rev. Lett.},\ }\textbf {\bibinfo {volume} {71}},\
      \bibinfo {pages} {1051} (\bibinfo {year} {1993})}\BibitemShut {NoStop}%
    \bibitem [{\citenamefont {Filippetti}\ and\ \citenamefont
      {Fiorentini}(1999)}]{Filippetti1999}%
      \BibitemOpen
      \bibfield  {author} {\bibinfo {author} {\bibfnamefont {A.}~\bibnamefont
      {Filippetti}}\ and\ \bibinfo {author} {\bibfnamefont {V.}~\bibnamefont
      {Fiorentini}},\ }\Doi {10.1103/PhysRevB.60.14366} {\bibfield  {journal}
      {\bibinfo  {journal} {Phys. Rev. B},\ }\textbf {\bibinfo {volume} {60}},\
      \bibinfo {pages} {14366} (\bibinfo {year} {1999})}\BibitemShut {NoStop}%
    \bibitem [{\citenamefont {Mura}\ \emph {et~al.}(2003)\citenamefont {Mura},
      \citenamefont {Ruggerone},\ and\ \citenamefont {Fiorentini}}]{Mura2003}%
      \BibitemOpen
      \bibfield  {author} {\bibinfo {author} {\bibfnamefont {M.}~\bibnamefont
      {Mura}}, \bibinfo {author} {\bibfnamefont {P.}~\bibnamefont {Ruggerone}}, \
      and\ \bibinfo {author} {\bibfnamefont {V.}~\bibnamefont {Fiorentini}},\ }\Doi
      {10.1103/PhysRevB.67.153406} {\bibfield  {journal} {\bibinfo  {journal}
      {Phys. Rev. B},\ }\textbf {\bibinfo {volume} {67}},\ \bibinfo {pages}
      {153406} (\bibinfo {year} {2003})}\BibitemShut {NoStop}%
    \bibitem [{\citenamefont {Pyykk\"o}\ and\ \citenamefont
      {Desclaux}(1979)}]{Pyykko1979}%
      \BibitemOpen
      \bibfield  {author} {\bibinfo {author} {\bibfnamefont {P.}~\bibnamefont
      {Pyykk\"o}}\ and\ \bibinfo {author} {\bibfnamefont {J.~P.}\ \bibnamefont
      {Desclaux}},\ }\href {http://dx.doi.org/10.1021/ar50140a002} {\bibfield
      {journal} {\bibinfo  {journal} {Acc. Chem. Res.},\ }\textbf {\bibinfo
      {volume} {12}},\ \bibinfo {pages} {276} (\bibinfo {year} {1979})}\BibitemShut
      {NoStop}%
    \bibitem [{\citenamefont {Pyykk\"o}(1988)}]{Pyykko1988}%
      \BibitemOpen
      \bibfield  {author} {\bibinfo {author} {\bibfnamefont {P.}~\bibnamefont
      {Pyykk\"o}},\ }\href {http://dx.doi.org/10.1021/cr00085a006} {\bibfield
      {journal} {\bibinfo  {journal} {Chem. Rev.},\ }\textbf {\bibinfo {volume}
      {88}},\ \bibinfo {pages} {563} (\bibinfo {year} {1988})}\BibitemShut
      {NoStop}%
    \bibitem [{\citenamefont {Christensen}(1984)}]{Christensen1984}%
      \BibitemOpen
      \bibfield  {author} {\bibinfo {author} {\bibfnamefont {N.~E.}\ \bibnamefont
      {Christensen}},\ }\Doi {10.1002/qua.560250119} {\bibfield  {journal}
      {\bibinfo  {journal} {Int. J. Quant. Chem.},\ }\textbf {\bibinfo {volume}
      {25}},\ \bibinfo {pages} {233} (\bibinfo {year} {1984})}\BibitemShut
      {NoStop}%
    \bibitem [{\citenamefont {Gambardella}\ \emph {et~al.}(2003)\citenamefont
      {Gambardella}, \citenamefont {Rusponi}, \citenamefont {Veronese},
      \citenamefont {Dhesi}, \citenamefont {Grazioli}, \citenamefont {Dallmeyer},
      \citenamefont {Cabria}, \citenamefont {Zeller}, \citenamefont {Dederichs},
      \citenamefont {Kern}, \citenamefont {Carbone},\ and\ \citenamefont
      {Brune}}]{Gambardella2003}%
      \BibitemOpen
      \bibfield  {author} {\bibinfo {author} {\bibfnamefont {P.}~\bibnamefont
      {Gambardella}}, \bibinfo {author} {\bibfnamefont {S.}~\bibnamefont
      {Rusponi}}, \bibinfo {author} {\bibfnamefont {M.}~\bibnamefont {Veronese}},
      \bibinfo {author} {\bibfnamefont {S.~S.}\ \bibnamefont {Dhesi}}, \bibinfo
      {author} {\bibfnamefont {C.}~\bibnamefont {Grazioli}}, \bibinfo {author}
      {\bibfnamefont {A.}~\bibnamefont {Dallmeyer}}, \bibinfo {author}
      {\bibfnamefont {I.}~\bibnamefont {Cabria}}, \bibinfo {author} {\bibfnamefont
      {R.}~\bibnamefont {Zeller}}, \bibinfo {author} {\bibfnamefont {P.~H.}\
      \bibnamefont {Dederichs}}, \bibinfo {author} {\bibfnamefont {K.}~\bibnamefont
      {Kern}}, \bibinfo {author} {\bibfnamefont {C.}~\bibnamefont {Carbone}}, \
      and\ \bibinfo {author} {\bibfnamefont {H.}~\bibnamefont {Brune}},\ }\Doi
      {10.1126/science.1082857} {\bibfield  {journal} {\bibinfo  {journal}
      {Science},\ }\textbf {\bibinfo {volume} {300}},\ \bibinfo {pages} {1130}
      (\bibinfo {year} {2003})}\BibitemShut {NoStop}%
    \bibitem [{\citenamefont {H\"akkinen}\ \emph {et~al.}(2002)\citenamefont
      {H\"akkinen}, \citenamefont {Moseler},\ and\ \citenamefont
      {Landman}}]{Hakkinen2002}%
      \BibitemOpen
      \bibfield  {author} {\bibinfo {author} {\bibfnamefont {H.}~\bibnamefont
      {H\"akkinen}}, \bibinfo {author} {\bibfnamefont {M.}~\bibnamefont {Moseler}},
      \ and\ \bibinfo {author} {\bibfnamefont {U.}~\bibnamefont {Landman}},\ }\Doi
      {10.1103/PhysRevLett.89.033401} {\bibfield  {journal} {\bibinfo  {journal}
      {Phys. Rev. Lett.},\ }\textbf {\bibinfo {volume} {89}},\ \bibinfo {pages}
      {033401} (\bibinfo {year} {2002})}\BibitemShut {NoStop}%
    \bibitem [{\citenamefont {Mavropoulos}(2003)}]{Mavropoulos2003}%
      \BibitemOpen
      \bibfield  {author} {\bibinfo {author} {\bibfnamefont {P.}~\bibnamefont
      {Mavropoulos}},\ }\href {http://stacks.iop.org/0953-8984/15/i=47/a=014}
      {\bibfield  {journal} {\bibinfo  {journal} {J. Phys. Condens. Matter},\
      }\textbf {\bibinfo {volume} {15}},\ \bibinfo {pages} {8115} (\bibinfo {year}
      {2003})}\BibitemShut {NoStop}%
    \bibitem [{\citenamefont {Pauling}(1929)}]{Pauling1929}%
      \BibitemOpen
      \bibfield  {author} {\bibinfo {author} {\bibfnamefont {L.}~\bibnamefont
      {Pauling}},\ }\href@noop {} {\bibfield  {journal} {\bibinfo  {journal} {J.
      Am. Chem. Soc.},\ }\textbf {\bibinfo {volume} {51}},\ \bibinfo {pages} {1010}
      (\bibinfo {year} {1929})}\BibitemShut {NoStop}%
    \end{thebibliography}
\end{document}